\magnification 1200

   \def\ni{\noindent}

\def\ie{{\it i.e.}}
\def\eq#1{ \eqno({#1}) \qquad }
\newbox\Ancha
\def\gros#1{{\setbox\Ancha=\hbox{$#1$}
   \kern-.025em\copy\Ancha\kern-\wd\Ancha
   \kern.05em\copy\Ancha\kern-\wd\Ancha
   \kern-.025em\raise.0433em\box\Ancha}}
%
%
%
\font\bigggfnt=cmr10 scaled \magstep 3
\font\biggfnt=cmr10 scaled \magstep 2
\font\bigfnt=cmr10 scaled \magstep 1
%
\vglue .75 in
\leftskip .25 in
\rightskip .25 in
\def\Par{\par\vskip 4 pt}
\baselineskip 16 pt

\centerline{\bigggfnt Jacobi equations using a variational principle}\Par

\vskip 18 pt
\centerline{\bigfnt H.\ N.\  N\'u\~nez-Y\'epez}\par

\centerline{Departamento de F\'{\i}sica}\par
\centerline{Universidad Aut\'onoma Metropolitana-Iztapalapa}\par
\centerline{ Apartado Postal  55-534 Iztapalapa 09340 D.\ F., M\'exico}\par
\centerline{e-mail: nyhn@xanum.uam.mx}
\vskip 12 pt

\centerline{\bigfnt   A.\ L.\ Salas-Brito\footnote{*}{\rm Corresponding author}}\par
\centerline{Laboratorio de Sistemas Din\'amicos}
\centerline{Departamento de Ciencias B\'asicas}\par
\centerline{Universidad Aut\'onoma Metropolitana-Azcapotzalco}\par
\centerline{ Apartado Postal 21-726, Coyoac\'an 04000 D.\ F., M\'exico}\par
\centerline{e-mail: asb@correo.azc.uam.mx}\Par
\vskip 12 pt

\centerline {\biggfnt Abstract} \par
 A variational principle is proposed for obtaining the Jacobi  equations in
systems admitting a Lagrangian description. The variational principle gives
simultaneously  the Lagrange equations of motion and the Jacobi variational
equations for the system. The approach can be of help in finding  constants of motion
in the Jacobi  equations as well as in analysing the  stability of the systems and can be related to the vertical extension of the Lagrangian 
formalism. To exemplify two of such aspects, we uncover a constant of motion in the Jacobi equations  of autonomous systems and we recover the well-known sufficient conditions of stability of two dimensional orbits in classical mechanics.

\vfill 
\noindent Classification numbers: 03.20.+i; 02.30.Wd; 02.40.Hw. \Par 

\eject 

Many classical dynamical systems have a variational formulation, for example, conservative mechanical systems, geodesic flows, classical field theory and even geometrical optics. All of them can be described using a  Lagrangian function  and a variant of  Hamilton's principle [1]. Given such common factors we call them Lagrangian dynamical systems. The variational formulation  of such Lagrangian systems is not only elegant and compact it also  allows uncovering deep connections between  the dynamical properties and other fields of physics and mathematics [1--6]. It is the goal of this short communication to formulate a generalized variational principle for Lagrangian dynamical systems, similar to Hamilton's, which is capable of producing not only the usual equations of motion but also the dynamical equations for deviations between two nearby trajectories: the so-called  Jacobi variational  equations [1,2]. 
 \Par 

We   discuss a variational principle  for what is  known as the equation of geodesic deviations in gravitational physics and in Riemannian geometry [2,4,7--8] and give a  brief outline of the possible applications of its consequences. Let us  pinpoint that the Jacobi  equations can be also regarded as the basic equations for deciding questions of dynamical stability, for evaluating  the Liapunov spectrum in dynamical systems [10--11],  that  they can be shown to occur naturally in the framework of the  vertical extension of the Lagrangian formalism [3], and, possibly, may be used to recast the Jacobi field generated geoodular structure of affinely connected manifolds [4]. \Par

Let us begin formulating  the variational principle. Let us assume a $N$-degree of freedom system that can be 
described by a Lagrangian $ L({\bf q}, {\bf \dot q}, t)$, where ${\bf q}$ and ${\bf \dot
q}$ stand, respectively,  for the $N$ generalized coordinates $(q_1, q_2,\dots q_n)$ and the $N$ generalized velocities $(\dot
q_1, \dot q_2,\dots \dot q_n)$. Using the Lagrangian of the system,  define the function $\gamma({\bf q}, {\bf\dot q}, {\gros \epsilon}, \dot {\gros\epsilon}, t)$ as

$$ \gamma({\bf q}, {\bf\dot q}, {\gros \epsilon}, \dot {\gros\epsilon}, t)\equiv 
{\partial L\over \partial {\dot q}_a} \dot \epsilon_a + {\partial L\over \partial 
q_a} \epsilon_a \eq{1} $$

\noindent  here, as in the rest of the article,  the summation convention (summing from 1 to $N$) is implied for repeated indices. The  $N$-vectors  ${\gros \epsilon}=(\epsilon_1,\epsilon_2,\dots, \epsilon_N)$ and $ {\bf\dot{\gros \epsilon}}=(\dot\epsilon_1,\dot\epsilon_2,\dots, \dot\epsilon_N)$ are to be regarded as describing deviations, and their corresponding velocities, from the motion  described by ${\bf q}$ and ${\bf \dot q}$. That is, ${\gros\epsilon}$ plays the role of the Jacobi field associated with the trajectories of the original system [1,2,4]. Let us  notice the important property that $\gamma({\bf q}, {\bf \dot q},{\gros \epsilon}, \dot {\gros\epsilon}, t)$ is an explicit function of time {\sl only} when the Lagrangian  is time-dependent (\ie\ when it is non-autonomous).\Par

Given the function $\gamma$ defined in (1), let us introduce the functional

$$  \Sigma[{\bf q}(t), {\gros\epsilon}(t)]=\int_{t_1}^{t_2} \gamma({\bf q}, {\bf \dot q}, {\gros\epsilon}, {\bf \dot \gros\epsilon},t)  dt \eq{2}$$

\noindent of the paths joining two given configurations $(q_1,\dot q_1,\epsilon_1, \dot \epsilon_1)$ and 
$(q_2,\dot q_2,\epsilon_2, \dot \epsilon_2)$ of the varied system between two
instants of time $t_1$ and $t_2$. The statement of the variational principle is now just that

$$ \delta \Sigma[{\bf q}(t), {\gros\epsilon}(t)]=0. \eq{3}  $$

\ni when the path is varied  with the endpoints and the time fixed. The $2N$
conditions guaranteeing that the functional $\Sigma[q(t), \epsilon(t)]$ takes an extremum value are the associated Euler-Lagrange equations [1]

$$ \eqalign{    &{d\over dt }\left( {\partial \gamma\over \partial \dot \epsilon_a} \right)
-{\partial \gamma\over \partial \epsilon_a}=0,\cr
&{d\over dt }\left( {\partial \gamma\over \partial \dot q_a} \right)-
{\partial \gamma\over \partial q_a}=0,\; a=1,\dots,N;} \eq{4}$$

\ni or, using the definition (1) in the preceding equations, we obtain the $N$ Lagrage equations of the original system

$$ {d\over dt }\left( {\partial L\over \partial \dot q_a} \right)-
{\partial L\over \partial q_a}=0,\quad a=1,\dots, N, \eq{5}$$

\ni plus the $N$   equations:

$$    M_{ab}\ddot \epsilon_b + C_{ab}\dot \epsilon_b  +
K_{ab}\epsilon_b=0,   \quad a =1,\dots, N, \eq{6}                 $$

\ni for the deviation, ${\gros \epsilon}$, between two nearby trajectories. The $N\times N$ matrices $M$, $C$ and $K$, are defined by

$$ \eqalign{M_{ab}&=\left({\partial^2 L\over \partial \dot q_a \partial \dot
q_b}\right), \quad C_{ab}=\left[{d\over dt}\left( {\partial^2L\over \dot q_a\dot q_b}
\right) + {\partial^2 L\over \partial \dot q_a \partial q_b} -{\partial^2 L\over
\partial \dot q_b \partial q_a}\right], \cr
K_{ab}&=\left[{d\over
dt}\left( {\partial^2L\over \dot q_a q_b} \right)  -{\partial^2 L\over \partial 
q_a \partial q_b}\right], \quad a,b=1,\dots,N.} \eq{7}$$

\ni Equations (6) are the Jacobi variational equations for the original system [1,2,4].\Par

If the Lagrangian of the system is time-independent, the  system has a well-known constant of motion

$$ H= {\partial L\over \partial \dot q_b}\dot q_b -L; \eq{8} $$

\ni in such time-independent case the Jacobi equations also admit---hence the importance of the property of $\gamma$ mentioned above--- an analogous
constant, namely

$$ h= {\partial \gamma \over \partial \dot q_b}\dot q_b +
{\partial \gamma\over \partial \dot \epsilon_b}\dot \epsilon_b; \eq{9} $$

\ni using definition (1), the constant $h$ can be cast in the form

$$ h= {\partial H\over \partial \dot q_b}\dot \epsilon_b +
{\partial H\over \partial  q_b} \epsilon_b. \eq{10} $$\Par

\ni Equation (9) [or (10)] is an important conclusion from the variational formulation.

 Notice that the variational equations can be written in  first-order form provided that

$$ \det\left| {\partial^2 L\over \partial \dot q_a \partial \dot q_b} \right|\neq 0,
\eq{11} $$

\ni \ie\ that the matrix $M$  is invertible; this is also the condition for the existence of a Hamiltonian description of the system [1]. Granted such condition, the $2N$ variational equations can be written as

$$ {d{\bf x}\over dt}= {\cal J}\cdot{\bf x},   \eq{12} $$

\ni where the $2N$-vector ${\bf x}$ and the $2N \times 2N$ matrix $\cal J$, are defined respectively by

  $$ {\bf x}=\pmatrix{{\gros \epsilon}\cr
                          {\bf \dot \gros\epsilon} },  \qquad {\cal J}=\pmatrix{0 &1 \cr
                                                              M^{-1}K &M^{-1}C},
\eq{13}$$

\ni where the 0 and the 1 stand,  respectively, for $N\times N$ zero and unit
matrices.\Par

We can now, for example,  use the solutions to equations (12) [or to (5) and (6)] to evaluate the $N$ Liapunov exponents, $\lambda_a$, in the standard way [8,11]

$$ \lambda_a=\lim_{t\to \infty} {1\over t}\log {||{\bf x}_a(t)||\over ||{\bf
x}_a(0)||},\quad a=1,\dots, N.\eq{14} $$

\ni where $||\bf v ||$ is any norm of the $N$-vector $\bf v$ and ${\bf x}(0)$ is just the initial condition for a perturbation, in one of the $N$ appropriate directions, of the orbit ${\bf q}(t)$ under analysis.\Par

 We emphasize that the variational formulation of the Jacobi equations can be of help for discovering constants of motion in the variational equations of particular systems (using their symmetries, for example), like the one we derived here [equation (9)] for the autonomous case. The close relationship of the variational principle to the Lagrangian through definition (1) reflects the central relationship between $L({\bf q}, {\bf \dot q}, t)$ and the properties of the motion. The Lagrangian is thus, paraphrasing an  apt description, the true gene of the motion [12].\Par

  The results derived from our variational principle  have some bearings on matters of stability. For example, using our results, we can analyse  the stability of particle orbits  in two dimensions recovering  well-stablished results in the process. \Par

To illustrate the previous assertion, let us analyse the motion of a particle with unit mass under a particular time independent potential, $U({\bf q})$, in two dimensions [9,13].  Select  a particular  orbit ${\cal O}$ as a reference,  the generalized coordinates can now be chosen as  the distance, $z$, from the reference orbit to the particle and the arc length, $s$,  from an arbitrary origin on ${\cal O}$  to the point, also on the reference orbit ${\cal O}$, from which $z$ is measured.  The Lagrangian of the system and the function $\gamma$ in these coordinates are (in obvious notation)

$$ \eqalign{L(s, \dot s, z, \dot z)&= {1\over 2}\left(\dot z^2 + \dot s^2\left(1+ {z\over  \rho}\right)^2\right)- U(z, s),\cr
\gamma(\epsilon_s, \dot\epsilon_s, \epsilon_z, \dot \epsilon_z)&=\dot z \dot\epsilon_z+ \dot s\dot\epsilon_s\left(1+{z\over \rho}\right)^2 + \left[ {\dot s^2\over  \rho^2}(\rho+{z}) -{\partial U\over \partial z}\right]\epsilon_z -\cr
 &\left[{z\dot s^2\over \rho^3} (\rho+z){d\rho\over ds} + {\partial U\over \partial s}\right] \epsilon_s.} \eq{16}$$

\ni where $\rho$ is the radius of curvature of the reference orbit $ {\cal O}$ at $s$. The system is autonomous, hence the quantity

$$ h= \dot s \dot z + \left(\dot {s^2\over \rho} +{\partial U\over \partial z}\right) \epsilon_z + {\partial U\over \partial s} \epsilon_s,
            \eq{17}$$

\ni is a constant of the motion. Combining  (16) and (17), using  equations (5) and (6), and choosing  $\cal O$ as the orbit under analysis, we can show that the  equation governing the stability of that orbit is 

$$
\ddot \epsilon_z +{3\over \rho} \left( {\dot s^2\over \rho} + {\rho\over 3} {\partial^2 U\over \partial z^2}\right) \epsilon_z = {2 h\over \rho}, \eq{18} $$

\ni since the $\epsilon_s$ equation  concerns itself only with deviations tangent to the original orbit. The stability of the two-dimensional orbits is easily established in the $h=0$ case---which simply corresponds to analysing varied trajectories in which the energy does not change respect to its value in the nonvaried orbit $\cal O$. In this  homogeneous case, a well known result [14] asserts that the above equation has oscillatory solutions (hence, the analysed orbit is stable)  if the quantity between parenthesis  in (18) ---called the coefficient of stability--- is positive definite at every point on the orbit [13]. 

For the relationship to the stability of periodic orbits see, for example, [13, 15]. Furthermore, we think the variational principle (3) can be useful for explaining in  non-standard way some results concerning the relationship between singularities in the exponential map and the corresponding Jacobi fields in geodesic flows on manifolds [2] with its natural geoodular structures [4], and can have relevance in the study of both Lagrangian and
 time-dependent Hamiltonian mechanics and field theories in the context of their vertical extensions [3].

\Par

\vskip 4 pt
\ni{\bf Acknowledgements}\par
\vskip 6pt
\ni This work was partially supported by CONACyT (Mexico). We also acknowledge useful discussions with L.\ S.\ Micha and E.\ Ula. We also acknowledge with thanks  the comment and references provided by  G.\ Sardanashvily of Moscow State University and the reprint of Ref.\ 4 made available by A.\ Klimov of the Departamento de F\'{\i}sica, Universidad de Guadalajara.
 ALSB wants to thank the Department of Physics of Emory University and, particularly, Professor F.\ Family for a stimulating one-month stay in Atlanta which provided, among other things, a certain amount of time to think about this subject. Last but not least, the authors dedicate this work to the memory of Ch.\ Cori.
\vfill
\eject

\ni {\bf References.}\Par

\ni\item {[1]}V.\ I.\ Arnold, {\it Mathematical Methods of Classical Mechanics}, Springer, New York, 1978.

\ni\item{[2]} M.\ P.\ Do Carmo, {\it Riemannian Geometry}, Birkh\"auser, Boston, 1992, Ch. 5. \par

\ni\item{[3]} G.\ Giachetta, L.\ Mangiarotti and G.\ Sardanashvily, J.\ Math.\ Phys., {\bf 40}, (1999) 1376; G.\ Giachetta, L.\ Mangiarotti and G.\ Sardanashvily, Differential
 geometry of time-dependent mechanics, arXiv: dg-ga/9702020 (Sections 4.7,
 5.5);  G.\ Giachetta, L.\ Mangiarotti and G.\ Sardanashvily, {\it New Lagrangian and
Hamiltonian Methods in Field Theory},  World Scientific, Singapore, 1997, Section 4.9;
L.\ Mangiarotti and G.\ Sardanashvily, {\it Gauge Mechanics}, World Scientific, Singapore, 1998, Section
 5.11.

\ni\item{[4]}  A.\ I.\ Nesterov, Algebras Groups and Geometry, {\bf 15} (1998) 25.

\ni\item{[5]} R.\ P.\ Feynman, R.\ B.\ Leighton, and M.\ Sands,  {\it The Feynman Lectures on Physics},  Addison-Wesley, Reading MA, 1966, Vol.\ 2, Chap.\ 19.

\ni\item{[6]} I.\ C.\ Percival, Am.\ Inst.\ Phys.\ Conf.\ Proc.\ {\bf 57} (1979) 302; reprinted in R.\ S.\ Mackay and J.\ D.\ Meiss (compilators), {\it Hamiltonian Dynamical Systems}, Adam Hilger, Bristol, 1987, p.\ 367.

\ni\item{[7]} H.\ Stephani, {\it General Relativity}, Cambridge University P., Cambridge, 1982, section 1.3.

\ni\item{[8]} R.\ Carretero-Gonz\'alez, H.\ N.\ N\'u\~nez-Y\'epez, and A.\ L.\ Salas-Brito, Phys. Lett. A {\bf 188} (1994) 48.\par

\ni\item{[9]} A.\ L.\ Salas-Brito, Am.\ J.\ Phys.\ {\bf 52} (1984) 1012.\par

\ni\item{[10]} R.\ S.\ Mackay in S.\ Sarkar (editor) {\it Nonlinear Phenomena and Chaos}, Adam Hilger, Bristol, 1986, p.\ 254.\par

\ni\item{[11]} C.\ M.\ Arizmendi, R.\ Carretero-Gonz\'alez, H.\ N.\ N\'u\~nez-Y\'epez, and A.\ L.\ Salas-Brito, in {\it New Trends in Hamiltonian Systems \& Celestial Mechanics: Advanced Series in Nonlinear Dynamics} {Vol.\ 8}, Eds.\ E.\ Lacomba and J.\ Llibre, World Scientific, Singapore, 1996, p.\ 1.\par

\ni\item{[12]}  D.\ Oliver, {\it The Shaggy Steed of Physics}, Springer, New York, 1994,  p. 36.

\ni \item{[13]} S.\ Chandrasekhar, {\it Principles of Stellar Dynamics}, University of Chicago Press, Chicago, 1942, Appendix 2.

\ni \item {[14]} E.\ L.\ Ince, {\it Ordinary Differential Equations}, Longmans, London, 1927, Ch 10.

\ni\item{[15]} R.\ S.\ MacKay, J.\ D.\ Meiss, Phys.\ Lett.\ A {\bf 98} (1983) 92.

 \vfill
  \eject 
 \end